\documentclass[preprint2]{aastex}

\usepackage{amsmath}

\slugcomment{Draft version July 24, 2016}

\shorttitle{Flux Ratio of Raman OVI 6825 and 7082}
\shortauthors{Lee et al.}

\begin{document}


\title{A Monte Carlo Study of Flux Ratios of Raman
Scattered O~VI Features at 6825 \AA\ and 7082 \AA\ 
in Symbiotic Stars}


\author{Young-Min Lee,  Dae-Sub Lee,  Seok-Jun Chang, Jeong-Eun Heo, Hee-Won Lee}
\affil{Department of Physics and Astronomy, Sejong University, Seoul, 05006, Korea}

\author{Narae Hwang, Byeong-Gon Park\altaffilmark{1}, Ho-Gyu Lee }
\affil{Korea Astronomy and Space Science Institute,  Daejon, 34055, Korea}

\altaffiltext{1}{Korea University of Science and Technology (UST),  Daejon, 34113,  Korea}




\begin{abstract}
Symbiotic stars are regarded as wide binary systems consisting of a hot white 
dwarf and a mass losing giant. They exhibit unique spectral features 
at 6825 \AA\ and 7082 \AA, which are formed via Raman scattering of 
\ion{O}{6}$\lambda\lambda$ 1032 and 1038 with atomic hydrogen. 
We adopt a Monte Carlo technique to generate the same number of \ion{O}{6}$\lambda$1032 and $\lambda$1038 line 
photons and compute the flux ratio
$F(6825)/F(7082)$ of these Raman scattered \ion{O}{6} features formed  in  neutral regions 
with a simple geometric shape as a function of \ion{H}{1} column density $N_{HI}$.  In cylindrical 
and spherical neutral regions
with the \ion{O}{6} source embedded inside, the flux ratio $F(6825)/F(7082)$ shows an overall decrease from 3 to 1 
as $N_{HI}$ increases in the range $10^{22-24}{\rm\ cm^{-2}}$. In the cases of a slab geometry and other geometries with the \ion{O}{6} source outside the
\ion{H}{1} region, Rayleigh escape operates to lower the flux ratio considerably. For moderate
values of $N_{HI}\sim 10^{23}{\rm\ cm^{-2}}$ the flux ratio behaves in a complicated
way to exhibit a broad bump with a peak value of 3.5 in the case of a sphere geometry. 
We find that the
ratio of Raman conversion efficiencies of \ion{O}{6}$\lambda\lambda$1032, 1038 ranges from 0.8 to 3.5.
Our high resolution spectra of 'D' type HM~Sge and 'S' type AG~Dra obtained 
with the {\it Canada-France-Hawaii-Telescope} show that the flux ratio 
$F(6825)/F(7082)$
of AG~Dra is significantly smaller than that of HM~Sge, implying that 'S' type symbiotics are characterized by
higher $N_{HI}$ than 'D' type symbiotics.
\end{abstract}





\section{Introduction\label{sec:intro}}

Binary systems involving an accreting white dwarf are classified into
cataclysmic variables and symbiotic stars, which
constitute important candidates for Type Ia supernovae \citep[e.g.][]{mikolajewska12}. In cataclysmic
variables, the secondary red dwarf star filling the Roche lobe 
injects its material through the inner Lagrangian point leading to
formation of an accretion disk around the white dwarf primary
\citep[e.g.][]{warner95}. In contrast, symbiotic stars consist of a compact star, mostly a white dwarf,
and a giant star losing a large amount of mass in the
form of a slow stellar wind. Some fraction of the slow stellar wind from the giant companion
is gravitationally captured by the white dwarf \citep[e.g.][]{mikolajewska12}.

Activities associated with symbiotic stars include X-ray emission, erratic
variabilities, prominent emission lines and collimated outflows \citep[e.g.][]{angeloni12, zamanov15}. 
In particular, \cite{angeloni11} discovered
a huge jet in Sanduleak's star. It is highly controversial
whether an accretion disk is formed  in a stable fashion in symbiotic stars. Hydrodynamical
studies show the plausibility of an accretion disk and also some
symbiotic stars are known to exhibit optical flickering indicating
the presence of an accretion disk \citep[e.g.][]{sokoloski01, sokoloski10}. 

Symbiotic stars are classified into 'S' type and 'D' type based on the spectral energy distribution 
in the IR region, where 'D' type symbiotics show an IR excess indicative of a dust component 
with a temperature $\sim 10^3{\rm\ K}$ \citep[e.g.][]{whitelock87,angeloni10}. No such IR excess is apparent 
in 'S' type symbiotics.
The orbital periods of 'D' type symbiotics are poorly known and are presumed to exceed several 
decades, implying that the giant companion is separated from the white dwarf by tens of AU. 
This is in high contrast in that many 'S' type symbiotics are known to have orbital periods of several
hundreds of days, which points out that the giant companion is much closer to the white dwarf 
than their 'D' type counterparts.

A significant fraction of symbiotic stars  exhibit unique broad emission features
at 6825 \AA\ and 7082 \AA\ with a width $\sim 20-30{\rm\ \AA}$. These
mysterious spectral features were identified by Schmid (1989),
who proposed that they are formed
through Raman scattering of O~VI$\lambda\lambda$ 1032 and 1038
with atomic hydrogen. Raman scattering proceeds with an incident far UV photon blueward of Ly$\alpha$ on
a hydrogen atom in the ground state, which finally de-excites
into the $2s$ state.  
As opposed to Raman scattering, the elastic version is Rayleigh scattering.

For an incident \ion{O}{6} line photon, the energy difference 
of the $1s$ and $2s$ states of a hydrogen atom is responsible for the creation of
a Raman photon redward of H$\alpha$.
More explicitly, an incident far UV photon
with frequency $\nu_i$ can be 
Raman scattered to re-appear with frequency $\nu_0$ given by
\begin{equation}
\nu_0=\nu_i - \nu_{Ly\alpha}
\end{equation}
where $\nu_{Ly\alpha}$ is the frequency of Ly$\alpha$.
To a frequency range around $\nu_i$ denoted by $\Delta\nu_i$
corresponds a frequency range $\Delta\nu_o$ around $\nu_o$, where 
$\Delta\nu_o=\Delta\nu_i$. Therefore, we have
\begin{equation}
{\Delta\nu_i\over\nu_i}
=\left({\nu_o\over\nu_i}\right){\Delta\nu_o\over\nu_o},
\end{equation}
which explains the broad line widths
exhibited by Raman scattered features \citep{nussbaumer89}.

The scattering cross section and the branching ratio
of Raman scattering by atomic hydrogen are discussed in a number
of research works including \cite{schmid89}, \cite{nussbaumer89}
and \cite{saslow69}. Adopting the values of
1031.928\AA\ and 1037.618 \AA\ as center wavelengths of \ion{O}{6} resonance doublet \citep{moore79}, 
the cross sections for Rayleigh and Raman scattering are 
\begin{eqnarray}
\hskip-10pt \sigma_{Ray}(1032)=34.0\ \sigma_T, && \hskip-25pt \quad \sigma_{Ram}(1032) =7.5\ \sigma_T
\nonumber \\ 
\hskip-10pt \sigma_{Ray}(1038)=6.6\ \sigma_T, && \hskip-25pt \quad \sigma_{Ram}(1038) =2.5\ \sigma_T
\end{eqnarray}
at line centers of \ion{O}{6} \citep{lee97a}.
Here, $\sigma_T =0.665\times 10^{-24}{\rm\ cm^2}$ is the Thomson
scattering cross section.

\cite{harries96} presented their spectropolarimetric survey
of Raman O~VI 6825 and 7082 features in a large number of symbiotic stars. In their study, 
most Raman O~VI fluxes show strong linear polarization 
with a degree of polarization amounting to $\sim 10$ percent.
In addition, the line profiles of these Raman features are characterized by double or triple
peak structures. Another notable point is that the Raman \ion{O}{6} 6825 and 7082 features 
exhibit disparate profiles.
In particular, usually the blue part of Raman 7082 feature is
relatively more suppressed than its Raman 6825 counterpart.
The profile disparity for Raman O~VI features was also
noted by other researchers including Schmid (1999), who proposed that
the kinematics associated with the giant wind is mainly responsible
for the multiple peak structures. \cite{lee99}  advanced a view that the
\ion{O}{6} emission region forms a part of the accretion flow around the white dwarf resulting
in multiple peak profiles in Raman \ion{O}{6} features \citep[see also][]{heo15}. 

\cite{allen80} used spectrophotometric data of a few symbiotic stars to report that the Raman 
O~VI feature at 6825 \AA\ is about 4 times stronger than the 7082 feature on average. 
\cite{schmid99} presented their spectroscopic observations of several symbiotic
stars to show that the flux ratio of Raman O~VI at 6825 \AA\ and 7082 \AA\
ranges from around 2 to 7. In particular, 'D' type symbiotics such as RR~Telescopii 
and V1016~Cygni exhibit high flux ratios  whereas 'S' type symbiotics show
low flux ratios. They attributed to this tendency to higher Raman conversion
efficiency in `S' type symbiotics than in `D' type ones,
where the Raman conversion efficiency is defined as the number ratio
$N({\rm Ram})/N({\rm O\ VI})$ between the Raman scattered photons and the incident \ion{O}{6} 
line photons.

Considering the role played by the Raman \ion{O}{6} features as diagnostic
of the accretion flow in symbiotic stars, studies of fundamental
properties on the conversion efficiency are of importance for proper
interpretation of spectroscopic data.
A pioneering  and fairly comprehensive work using a Monte Carlo technique
was presented by \cite{schmid92}, who included angular distribution
and polarization in his simulations.  Basic scattering geometries such as
an illuminated slab and sphere were considered to reveal the systematic changes
in flux and polarization expected in a binary orbital motion.
\cite{schmid96} and \cite{lee97a} also presented more extended studies of Raman scattered \ion{O}{6}
in scattering geometries that include an expanding \ion{H}{1} region around the giant component.

Despite these sophisticated and extensive previous works,  we assert that useful information regarding a
representative \ion{H}{1} column density around the giant component can be gained through simple
Monte Carlo simulations, which we present in this paper. We focus on the dependence on the \ion{H}{1} column
density of the flux ratio of Raman features at  6825  \AA\  and  7082 \AA\  in a few simple scattering geometries.

This paper is composed as follows.  In Section 2 we describe the scattering
geometry and the Monte Carlo procedures. We present our results in Section 3.
In Section 4, we present our high resolution spectra 
of the two symbiotic stars
AG~Draconis and HM~Sagittae obtained with the {\it Canada-France-Hawaii Telescope}
to infer the characteristic \ion{H}{1} column density in these systems.
We conclude and discuss observational implications in the final section.

\begin{figure}
\centering
\includegraphics[scale=.30]{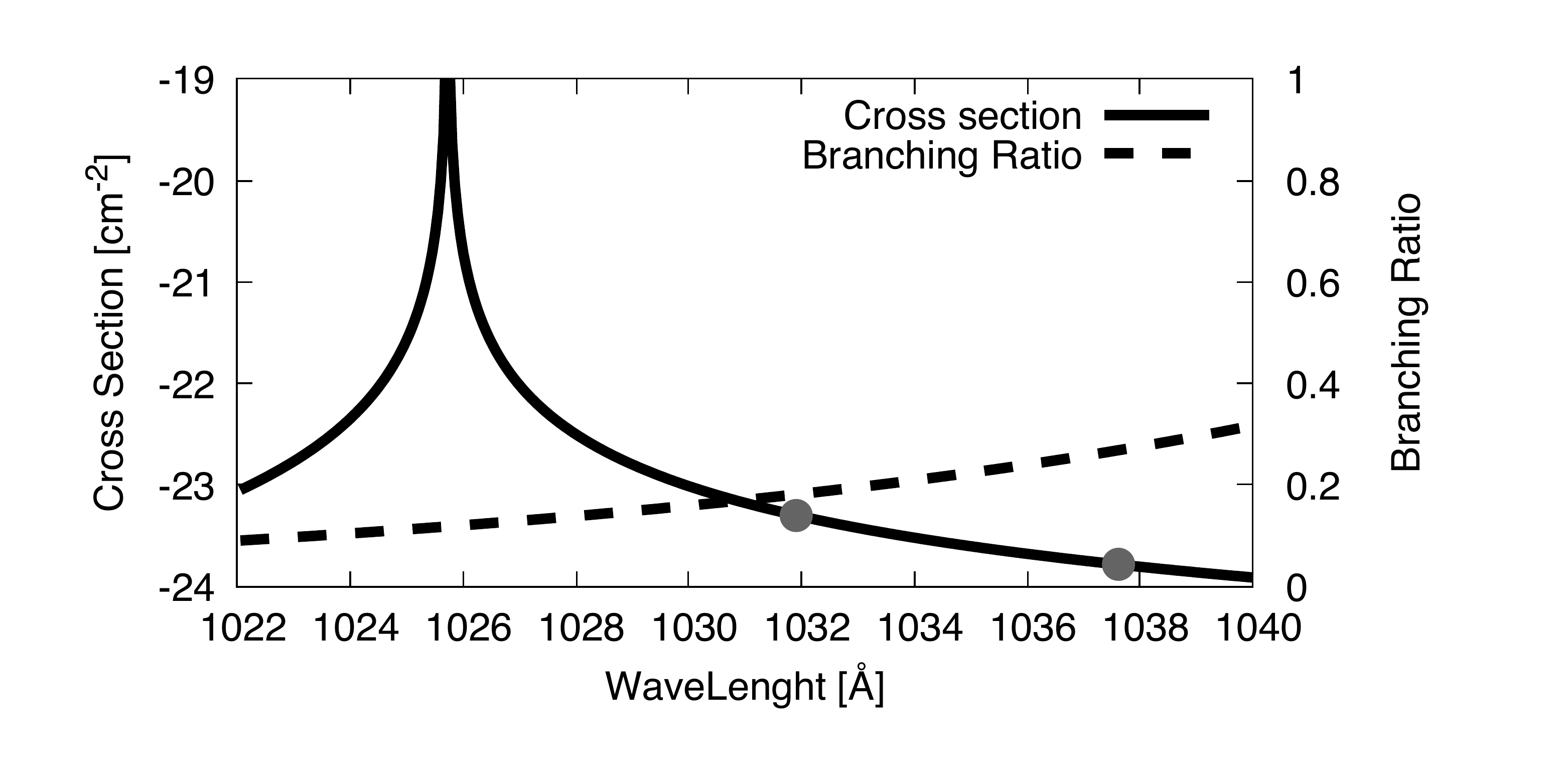}
\caption{Cross section and branching ratio for Raman scattering 
into $2s$ in a wavelength interval $1022{\rm\ \AA}<\lambda
<1040{\rm\ \AA}$. The left vertical axis shows the logarithm of the
cross section and the right vertical axis shows the branching ratio.
The branching ratio is an increasing function of
wavelength in this interval.
}
\end{figure}

\section{Monte Carlo Procedures}

\begin{figure*}
\centering
\includegraphics[scale=.30]{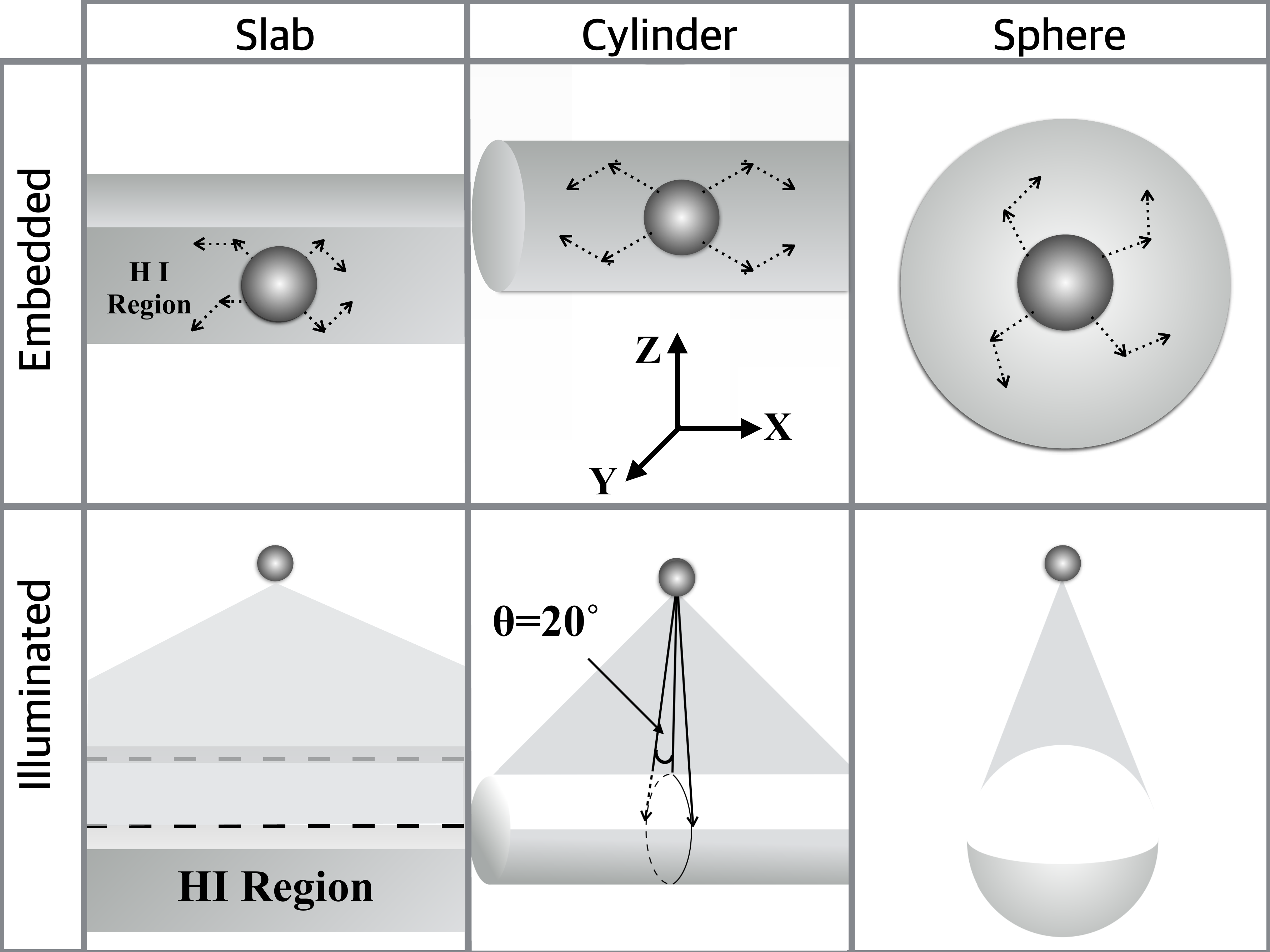}
\caption{Scattering geometry and the coordinate system considered in this work.
The neutral region takes a shape of a slab, a cylinder or a sphere. 
The top 3 panels correspond to the cases where the O~VI emission source is embedded at the center of the \ion{H}{1} region.
 In the bottom 3 panels,  the \ion{O}{6} emission source is outside the neutral region. The emission source is 
on the $z$-axis. In the cylinder and sphere cases, the circular radius subtends an angle of $20^\circ$ with respect to
the \ion{O}{6} emssion source. 
}
\label{scat_geometry}
\end{figure*}

In this work, we consider three types of geometric configuration for the neutral region, which are planar,
cylindrical and spherical. The neutral region is assumed to be static and of uniform \ion{H}{1} density $n_{HI}$.
The neutral slab considered in this work has a finite thickness
$H$ along the $z$-axis and is of infinite extent in the two lateral directions taken to be the $x-y$
directions. The cylinder has an infinite length along the $x$-axis and the circular section in the $y-z$ plane
is specified by the radius $R$. 
For each geometric shape, two cases are considered
for the location of a point-like and isotropic \ion{O}{6} emission source.
In the first case we embed  the \ion{O}{6} emission source at the
center of the neutral region. The second case is that the
\ion{O}{6} source is outside the neutral region. 

Fig.~\ref{scat_geometry} shows
schematically the scattering geometry and the coordinate system adopted in this work. The top 3 panels
correspond to the first case that the \ion{O}{6} source is located at the center of the neutral region.
The bottom 3 panels illustrate the second case where the neutral region is illuminated by the \ion{O}{6} source
located on the $z-$axis.
In  the cylinder case, the \ion{O}{6} source is located in such a way that the radius of the circular 
section cut by the $y-z$ plane subtends an angle 
of 20$^\circ$. Similarly in the case of the illuminated sphere, the radius $R$ subtends the same angle of $20^\circ$.

The scattering geometry is fully specified by determining the \ion{H}{1} column density $N_{HI}$ along
a designated direction. For example,  the slab geometry is specified by setting 
$N_{HI}=n_{HI}H$ along the
slab normal which coincides with the $z$-axis. In the cases of the cylinder and 
sphere geometry, we set $N_{HI}=n_{HI}R$. 

In this work, we also make a simple assumption that the neutral region is optically thin for optical
photons so that once Raman scattering takes place the scattered photon leaves the region to
without further interaction reach the observer. This assumption implies that in a neutral region
with a very high $N_{HI}$ the expected mean number of Rayleigh scatterings for an \ion{O}{6} line
photon is given by the inverse of the branching ratio, which is 5.5 for \ion{O}{6}$\lambda$1032
and 3.7 for \ion{O}{6}$\lambda$1038.

Our Monte Carlo simulation starts with a generation of an \ion{O}{6} line photon.
In this work, we assume that
the \ion{O}{6} emission  is isotropic and monochromatic at two frequencies corresponding to 
the line centers of \ion{O}{6}$\lambda\lambda$ 1032 and 1038. 
Two uniform random numbers $r_1, r_2$ in the interval $(0,1)$ are generated to obtain an isotropic unit wavevector
${\bf\hat k}$ associated with the initial photon
\begin{equation}
{\bf\hat k}=(\sin\theta\cos\phi,\sin\theta\sin\phi,\cos\theta)
\end{equation}
with $\theta=\cos^{-1}(2r_1-1)$ and $\phi=2\pi r_2$.

The physical distance $s$ traveled by an \ion{O}{6} line photon is related to the optical 
path length $\tau$ by
\begin{equation}
s=\tau/(n_{HI}\sigma_{tot}),
\end{equation}
where $\sigma_{tot}=\sigma_{Ray}+\sigma_{Ram}$  is the total scattering cross section of the \ion{O}{6} line photon. Subsequently, the next 
scattering site ${\bf r}_f$ from the initial position ${\bf r}_i$
is determined by
\begin{equation}
{\bf r}_f={\bf r}_i+s{\bf\hat k}.
\end{equation}
If ${\bf r}_f$ is inside the scattering region, a new scattering is made. Because
the same scattering phase function is shared by both Rayleigh and Raman scattering processes
\citep[e.g.][]{schmid90, lee97a},
we select a new unit wavevector before we determine the scattering type on the basis 
of the branching ratio.

A density matrix formalism is adopted to select the new unit wavevector, which is explained in the
literature \cite[e.g.,][]{chang15, ahn15}. Even though we disregard the polarization of the
emergent Raman scattered radiation in this work, each photon in our Monte Carlo simulation
is traced until escape and detection by an observer with polarization information retained.  
The density matrix associated with a photon in this simulation is a 2$\times$2 Hermitian matrix $\rho$
carrying the same information as the set of 4 Stokes parameters $I, Q, U$ and $V$. One representation
can be written as
\begin{equation}
\rho_{11}={I+Q\over2}, \quad \rho_{22}={I-Q\over2},
\quad 
\rho_{12}={U+iV\over 2}.
\end{equation}
Here, in this work no circular polarization is considered and $V=0$. 
The basis vectors for linear polarization are chosen in the following way
\begin{eqnarray}
{\bf\hat e}_1 &=& (-\sin\phi,\cos\phi,0)
\nonumber \\
{\bf\hat e}_2 &=& (\sin\theta\cos\phi,\sin\theta\sin\phi,\cos\theta),
\end{eqnarray}
where ${\bf\hat e}_1$ and ${\bf\hat e}_2$  indicate the polarization direction perpendicular and parallel 
to the $z-$axis, respectively. 

The unnormalized density matrix elements associated with the scattered radiation is given by
\begin{equation}
\rho'(\mu,\phi,\mu'\phi')_{kl}=\sum_{i,j=1}^2
|{\bf\hat e}'_k \cdot{\bf\hat e}_i| \rho_{ij}|{\bf\hat e}_j\cdot{\bf\hat e}'_l|.
\end{equation}
The unnormalized probability density function describing the angular distribution of the scattered radiation 
is obtained by taking the trace of the density matrix associated with the scattered radiation $\rho_{11}'+\rho_{22}'$,
which is explicitly given by
\begin{eqnarray}
Tr(\rho') &=& \rho'_{11}+\rho'_{22}
\nonumber \\
&=&\rho_{11}(\cos^2\Delta\phi+\mu'^2\sin^2\Delta\phi)
+\rho_{12}[-\mu\sin2\Delta\phi
\nonumber \\
&+&\mu'(\mu\mu'\sin2\Delta\phi
+2\sqrt{1-\mu^2}\sqrt{1-\mu'^2}\sin\Delta\phi)
\nonumber \\
&+ &
\rho_{22}[\mu^2\sin^2\Delta\phi +(\mu\mu'\cos\Delta\phi
\nonumber \\
&+&
\sqrt{1-\mu^2}\sqrt{1-\mu'^2})^2]
\end{eqnarray}

Typically we generate $10^6$ line photons for each data point. In this work, we do not consider the
angular distribution of Raman scattered radiation and collect all the emergent Raman photons.

\section{Result}

\subsection{Slab}

\begin{figure}
\centering
\includegraphics[scale=.22]{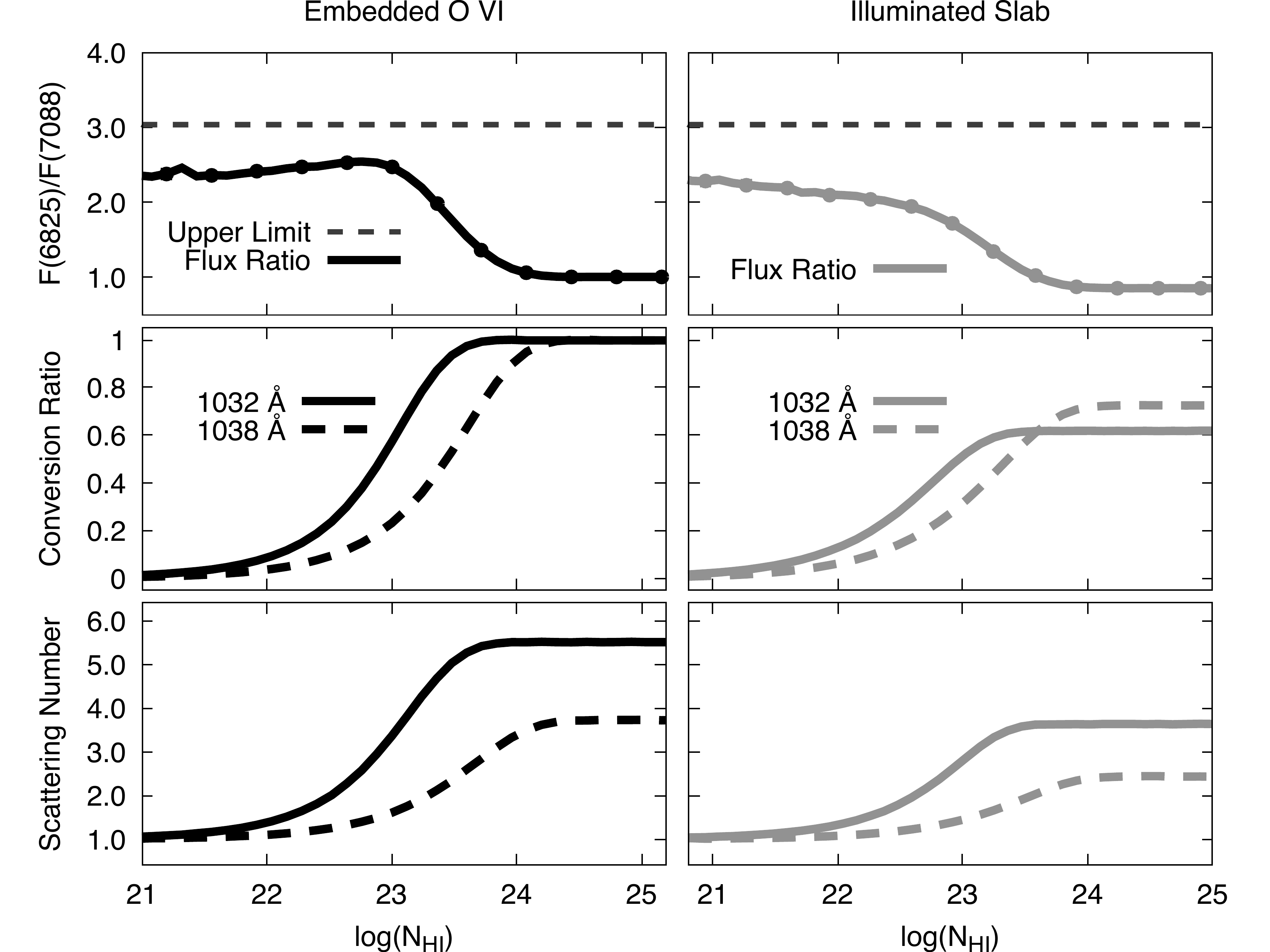}
\caption{Flux ratio, Raman conversion efficiency and mean scattering number of Raman scattered \ion{O}{6} 
emergent from a slab with a finite thickness
and infinite lateral dimensions. The left 3 panels and right 3 panels correspond to the cases where the \ion{O}{6} source
is embedded at the center of the slab and outside the slab, respectively.
The horizontal axis is the logarithm
of H~I column density measured along the slab normal. The top panels show the flux ratio
of Raman 6825 and Raman 7082.  The horizontal dotted line in each top panel
shows the ratio $\sigma_{Ram}(1032)/\sigma_{Ram}(1038)=3.04$.
The middle panels show the Raman conversion rate, and
the bottom panels show the mean number of scatterings before escape the neutral region.  
}
\label{slab_in_res}
\end{figure}

In Fig.~\ref{slab_in_res}, we show our Monte Carlo results for Raman scattered \ion{O}{6} 
formed in a neutral region with a slab geometry.
The left 3  panels correspond to the cases where the emission source is embedded at the center of the slab, and the right 3 panels are
for the case where the emission source is outside the slab.
The horizontal axis is the logarithm of
$N_{HI}=n_{HI}H$, where $H$ is the thickness of the slab. 

The top panels show the flux ratio $F(6825)/F(7082)$. The values obtained through Monte Carlo
calculations are shown by small circles that are connected by a solid curve.
The horizontal dashed line
indicates the ratio of Raman scattering cross sections for \ion{O}{6}$\lambda\lambda$ 1032 and 1038, which is 3.04.
This is the flux ratio naively expected in a very optically thin scattering region.
In the middle panels, we plot the Raman conversion efficiency, where the solid and dotted curves 
are for Raman \ion{O}{6} features at 6825 \AA\ and 7082 \AA, respectively.
The bottom panels show the mean scattering number. It should be noted that the minimum number of scatterings is
one for Raman scattered photons due to the inelastic nature of Raman scattering.

As $N_{HI}$ increases, the middle panels show that more and more fractions of \ion{O}{6} line photons are Raman converted 
to escape from the slab. In the case of the embedded \ion{O}{6} source, the conversion efficiency approaches unity.
The conversion efficiency in excess of 0.99 is reached for $N_{HI}\sim 10^{23.5}{\rm\ cm^{-2}}$ and $10^{24}{\rm\
cm^{-2}}$ for \ion{O}{6}$\lambda$1032 and 1038, respectively. 
 In this optically thick limit, the mean number of scatterings
before Raman conversion is given by the inverse of the branching ratio into Raman scattering. 
For \ion{O}{6}$\lambda$1032 and 1038, it is 5.5 and 3.7, respectively,
which is found in the bottom left panel in Fig.~\ref{slab_in_res}.

In the case where the \ion{O}{6} emission source is outside the \ion{H}{1} slab, as $N_{HI}$ increases, 
the Raman conversion rate converges to 0.62 and 0.72 for \ion{O}{6}$\lambda$1032 and 1038, respectively. 
 In this case a significant fraction $\sim 1/3$ of incident \ion{O}{6} photons escape from the slab via Rayleigh scattering
near the surface facing the \ion{O}{6} source.  This is also confirmed from the work of  
\cite{lee97b}, who presented
their result in terms of the parameter $R_{Ram} \sim 2$ defined as the number ratio of Raman photons and 
Rayleigh photons emergent from a slab with the total scattering optical depth $\tau_s =10$.  
This implies that roughly one third of incident \ion{O}{6} photons will escape through
Rayleigh scattering. They considered Raman scattering processes of a hypothetical UV photon
with the Raman branching ratio of 0.2, which is slightly larger than that for \ion{O}{6}$\lambda$1032 and 
smaller than that for \ion{O}{6}$\lambda$1038.

Rayleigh escape is more effective for \ion{O}{6}$\lambda$1032
than \ion{O}{6}$\lambda$1038, which is attributed to the larger Rayleigh branching ratio of \ion{O}{6}$\lambda$1032. 
Escape via Rayleigh scattering is significantly contributed by 
reflection events at first scattering site near the illuminated surface, which, in turn, is sensitive 
to the branching ratio. As is illustrated in the top right panel,
this Rayleigh reflection effect suppresses the formation of the Raman 6825 feature more than the Raman 7082 feature 
in this high $N_{HI}$ limit, which results in the flux ratio approaching a value $\sim 0.85$ smaller than unity.
A comparison of the bottom panels shows that the mean scattering number is lower in the illumination case, because
photons escaping from shallow regions of the illuminated surface are characterized by a small number of scatterings.

In the top panels, another interesting behavior in the flux ratio is noted in the optically thin limit, which is that
the flux ratio is not convergent to the
ratio of Raman scattering cross sections but remains significantly below it. It appears that the flux ratio converges to
a value $\sim 2.3$ for both cases. Because of the infinite extent in the $x-y$ directions, scattering optical 
depth for \ion{O}{6} photons traveling in the grazing direction can be significant,
leading to a mean number of scatterings slightly larger than 1, as shown in the bottom panels. 
While an \ion{O}{6} line photon can easily escape in the direction normal
to the slab, there is a nonnegligible fraction of \ion{O}{6} photons that escape from the thin slab through 
a single Rayleigh scattering. Escape through a single Rayleigh scattering can be made more readily 
by \ion{O}{6}$\lambda$1032 photons than \ion{O}{6}$\lambda$1038 ones due to larger branching ratio 
into Rayleigh scattering of \ion{O}{6}$\lambda$1032. This accounts for the flux ratio of 2.3 in the small 
$N_{HI}$ limit, which is smaller than the value of 3.05, the ratio of Raman scattering cross sections.

As is shown in the top right panel, the flux ratio monotonically decreases in the case where the \ion{O}{6} source 
is outside the slab. However, the flux ratio forms a broad hump peaking at $\sim 2.5$
near  $N_{HI}=10^{22.5}{\rm\ cm^{-2}}$ in the case where the \ion{O}{6} source is embedded in a neutral slab. 
 At this \ion{H}{1} column density, the total scattering optical depth
for \ion{O}{6}$\lambda$1032 is near unity, whereas \ion{O}{6}$\lambda$1038 remains Rayleigh-Raman optically thin.
A nonlinear behavior between the mean scattering number and $N_{HI}$ becomes significant for \ion{O}{6}$\lambda$1032,
while a linear relation prevails for \ion{O}{6}$\lambda$1038.


\subsection{Cylinder}

\begin{figure}
\centering
\includegraphics[scale=.22]{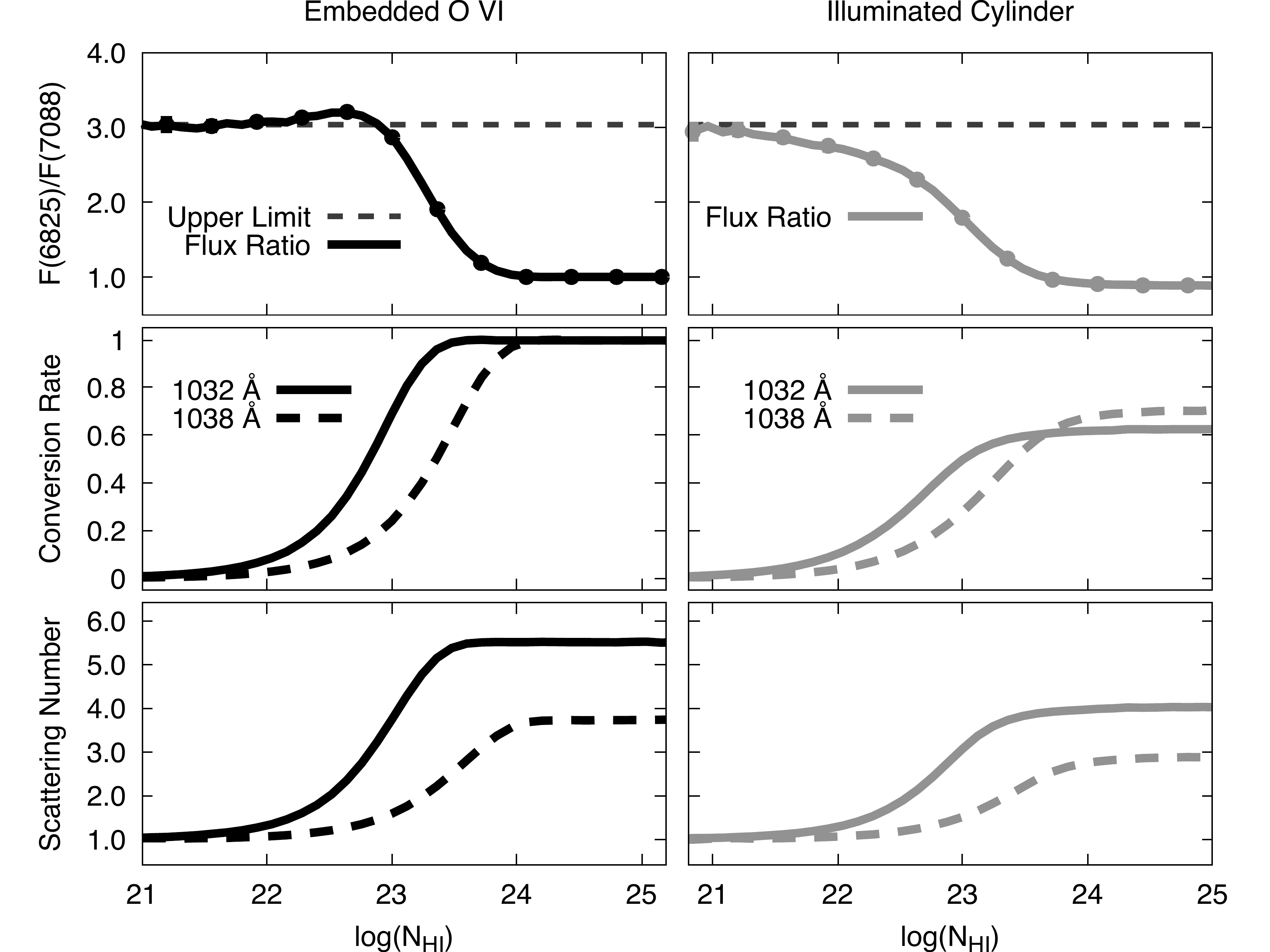}
\caption{The same quantities for a cylindrical neutral region as in Fig.~3.
The left 3 panels are for the results for the \ion{O}{6} emission source embedded at the center of the cylinder.
The right 3 panels are for the case where the \ion{O}{6} source is outside the cylinder with
the cylinder diameter subtending an angle of $20^\circ$ with respect to the \ion{O}{6} source. 
}
\label{cylinder_in_res}
\end{figure}

Fig.~\ref{cylinder_in_res} shows our Monte Carlo results for Raman scattered \ion{O}{6} formed 
in cylindrical neutral regions. The left 3 panels are for the case where the
\ion{O}{6} emission source is embedded at the center of the cylinder.
The flux ratio approaches the ratio of Raman scattering cross sections in the low $N_{HI}$ limit,
as is shown in the top left panel.
It is noted that escape through Rayleigh scattering is almost 
negligible in the cylindrical neutral region.
As in the case of the slab geometry with the emission source embedded, there is also a
broad hump near $N_{HI}\sim 10^{22.5}{\rm\ cm^{-2}}$ with the peak value of $\sim 3.2$.
As $N_{HI}$ increases, the flux ratio approaches unity, which implies the full conversion into
Raman optical photons. This is confirmed by the Raman conversion efficiency convergent to unity
shown in the left middle panel. The mean scattering number also converges to the inverse of
Raman branching ratio considered in the slab case.

In the right 3 panels,  we show our Monte Carlo result for a cylindrical neutral region
illuminated by an \ion{O}{6} emission source.  As shown in the top right panel, the flux
ratio monotonically decreases from the expected optically thin limit of 3.05 to a value $\sim 0.87$
as $N_{HI}$ increases. Again in this case 
Rayleigh reflection occurs near the boundary region facing the \ion{O}{6} source, which lowers the flux ratio 
to $\sim 0.87$ in the optically thick limit.
 This value is slightly larger than that obtained in the slab case
considered in Fig.~\ref{slab_in_res}. The Rayleigh reflection effect is contributed by
those photons incident in the grazing direction, of which the fraction is lower for a cylindrical geometry 
than for a slab geometry. This is also confirmed from the fact that the scattering numbers 
for \ion{O}{6}$\lambda$ 1032 and 1038 in the case of an illuminated cylinder
are larger than those values corresponding to the illuminated slab case.

\subsection{Sphere}

\begin{figure}
\centering
\includegraphics[scale=.22]{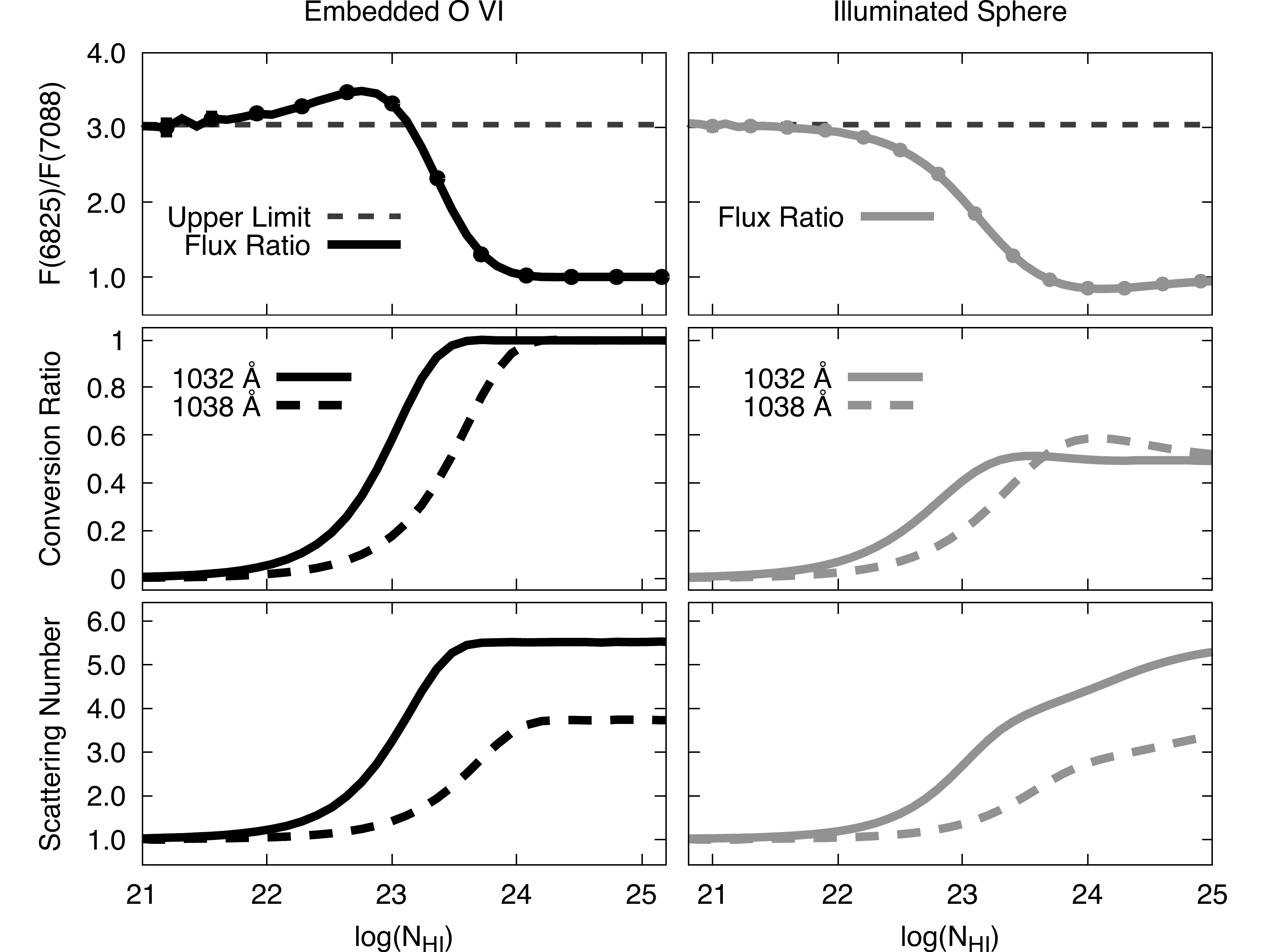}
\caption{The same quantities for a spherical neutral region as in Fig.~3.
The left 3 panels are for the results for the \ion{O}{6} emission source at the center of the sphere.
The right 3 panels are for the case where the \ion{O}{6} source is outside the sphere with
the diameter subtending an angle of $20^\circ$ with respect to the \ion{O}{6} source.  
}
\label{sph_in_res}
\end{figure}

Fig.~\ref{sph_in_res} shows our Monte Carlo result for Raman scattered \ion{O}{6} formed 
in a spherical neutral region. The left 3 panels are for the case where the emission source 
is embedded at the center. In the optically
thin limit, the flux ratio is convergent to the ratio of Raman scattering cross sections.
In the opposite case of an extremely high optical depth, the flux ratio approaches unity, 
which indicates the complete Raman conversion. We also note
that the mean number of scatterings also converges to the expected values, which are inverses
of the branching ratios.

In this figure, a particularly notable feature is the broad bump appearing at around $N_{HI}=10^{22.7}{\rm\
cm^{-2}}$. The peak value is approximately 3.5, a value significantly in excess of the ratio
of Raman scattering cross sections. The value of 3.5 is also the maximum flux ratio
obtained in this work.  Considering the fact that the flux ratio $F(1032)/F(1038)$
of \ion{O}{6}$\lambda$1032 and \ion{O}{6}$\lambda$1038 is 2 in an optically thin \ion{O}{6} nebular region, 
the largest value of the flux ratio $F(6825)/F(7082)$ that would be observed is about 7.

The right 3 panels show the Monte Carlo result for the case where the emission source is outside 
the sphere, illuminating it from a distance where the diameter of the H~I region subtends 
an angle of $20^\circ$.
In the optically thin limit, the flux ratio approaches the ratio of Raman scattering cross 
sections. In the limit of high optical depths the Rayleigh reflection effect is working so
that the flux ratio approaches 0.9, a value slightly less than unity.

It is also interesting to note the presence of a dip feature around $N_{HI}=10^{24}{\rm\
cm^{-2}}$ in the top right panel instead of the broad hump present in the top left panel in Fig.\ref{sph_in_res}. 
The broad hump  and dip features appearing in the top left and right panels of Fig.\ref{sph_in_res}, respectively, are 
attributed to the nonlinear behavior of the Raman
conversion efficiency as a function of $N_{HI}$ near unity of the Rayleigh-Raman scattering optical
depth for \ion{O}{6}$\lambda$1032 or \ion{O}{6}$\lambda$1038.

\subsection{Finite Cylinder Model}

\begin{figure}
\centering
\includegraphics[scale=.30]{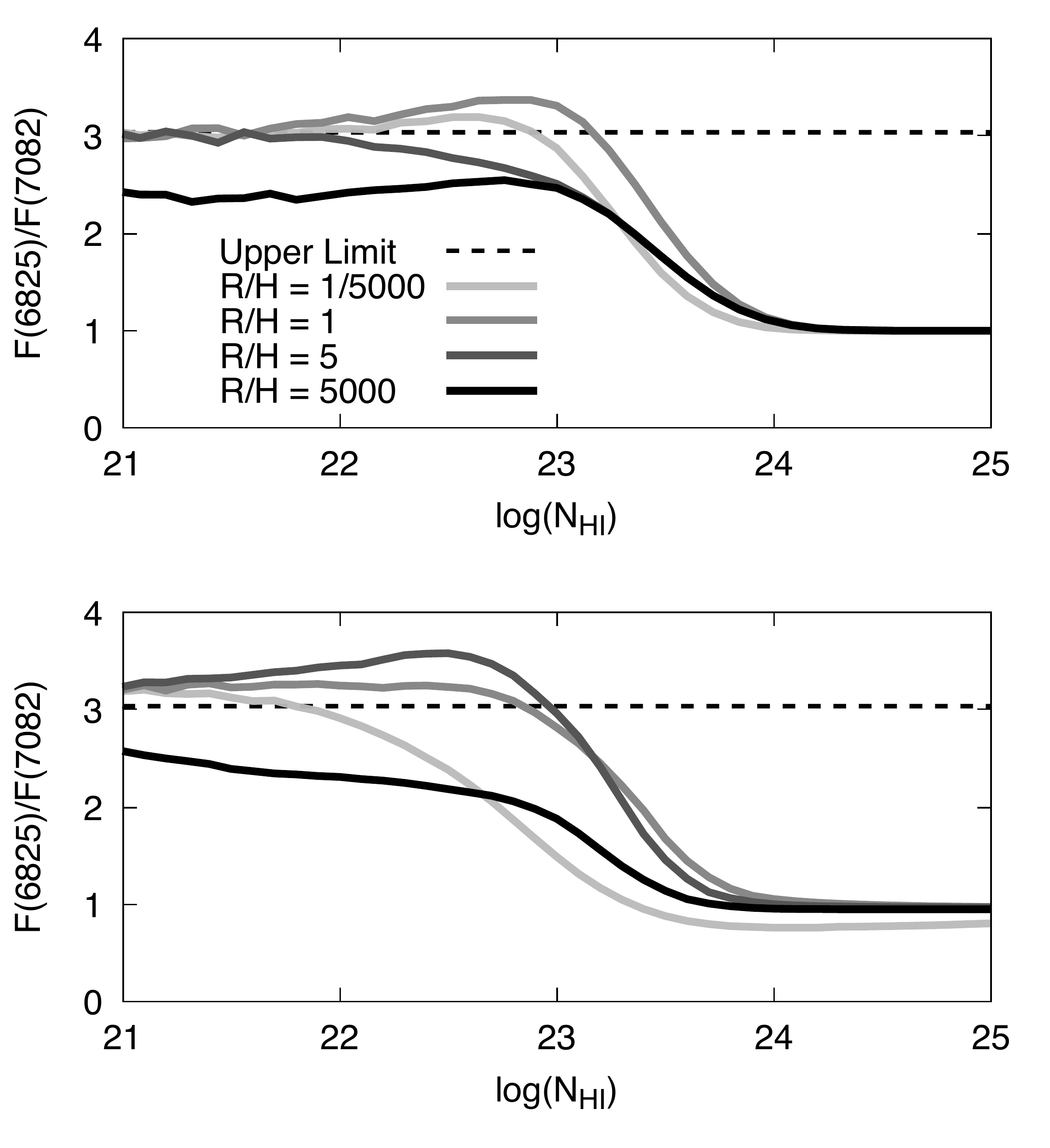}
\caption{Flux ratio $F(6825)/F(7082)$ of Raman scattered \ion{O}{6} emergent from a
 cylindrical neutral region. The cylinder has a radius $R$ and a height $H$. Four values of $R/H= 1/5,000, 1, 5$ and 5,000 are
considered. The emission source is located at the center of the cylinder in the top panel. In the bottom panel,
the \ion{O}{6} source is located on the extension of the cylinder axis illuminating the top circle at a distance where the
diameter subtends a right angle. The horizontal axis is the logarithm of
$N_{HI}$ along the cylinder axis so that $N_{HI}=n_{HI}H$. 
}
\label{fluxratios_res}
\end{figure}

In Fig.~\ref{fluxratios_res}, we present our result where the scattering region takes the form
of a cylinder with a finite height $H$ and a radius $R$. 
The horizontal axis is the logarithm of \ion{H}{1} column density $N_{HI}=n_{HI}H$ measured 
along the cylinder axis. 
We show the results for 4 values of $R/H=$   1/5000, 1, 5 and 5000. In the top panel, the \ion{O}{6} source 
lies at the center of the neutral region. In the bottom panel, the \ion{O}{6} source is located 
on the extension of the cylinder axis illuminating the top circle at a distance amounting to the radius $R$. 

The cases of $R/H=1/5000$ and 5000 are effectively the same as those of the cylinder and slab models, respectively, which we considered 
in the previous subsections. The case of $R/H=1$ also mimics the sphere model, in which the maximum
flux ratio of 3.5 is obtained. In these three cases a broad bump appears near $N_{HI} =10^{22.5}{\rm\ cm^{-2}}$.
However, as is shown in the case of $R/H=5$, the flux ratio monotonically decreases from the optically thin limit of 3
to the optically thick limit of 1.

 As $R/H$ increases from a very small value of $\sim 1/5000$ to a moderate value $\sim 1$, the change
in flux ratio occurs mainly in the transitional regime around $N_{HI}\sim 10^{23}{\rm\ cm^{-2}}$, where 
the flux ratio $F(6825)/F(7082)$ shows an overall increase. As $R/H$ further increases from  1 to  5,
a significant decrease in the flux ratio is made near $N_{HI}=10^{23}{\rm\ cm^{-2}}$, which results in an
overall monotonic decrease of $F(6825)/F(7082)$ in the entire range of $N_{HI}$. Now as $R/H$ increases 
sufficiently, Rayleigh escape lowers the flux ratio in the  regime $N_{HI}<10^{23}{\rm\ cm^{-2}}$
resulting in the overall flux ratio $\sim 2.3$ obtained in the slab case in the optically thin limit.

In the bottom panel there are a few interesting points to be noted. The first point is that when $R/H=5$ there appears a
broad bump at $N_{HI}=10^{22.6}{\rm\ cm^{-2}}$ with a peak value $\sim 3.5$. Considering a similar high value 
is obtained in the sphere case with an embedded \ion{O}{6} source, caution should be exercised to interpret
observed flux ratios of Raman \ion{O}{6} features.  Another point is that
when $R/H = 1$ the behavior is similar to that found for a neutral sphere illuminated outside considered in
Fig.~5. Finally as is found in the illuminated slab case, the effect of Rayleigh escape is clearly seen
in the case $R/H=1/5,000$, where $F(6825)/F(7082)$ remains $\sim 0.8$ for $N_{HI} \ge 10^{24}{\rm\ cm^{-2}}$.

\section{Comparison of Simulated and Observed Spectra}

\subsection{Simulated Spectra}

\begin{figure}
\centering
\includegraphics[scale=.24]{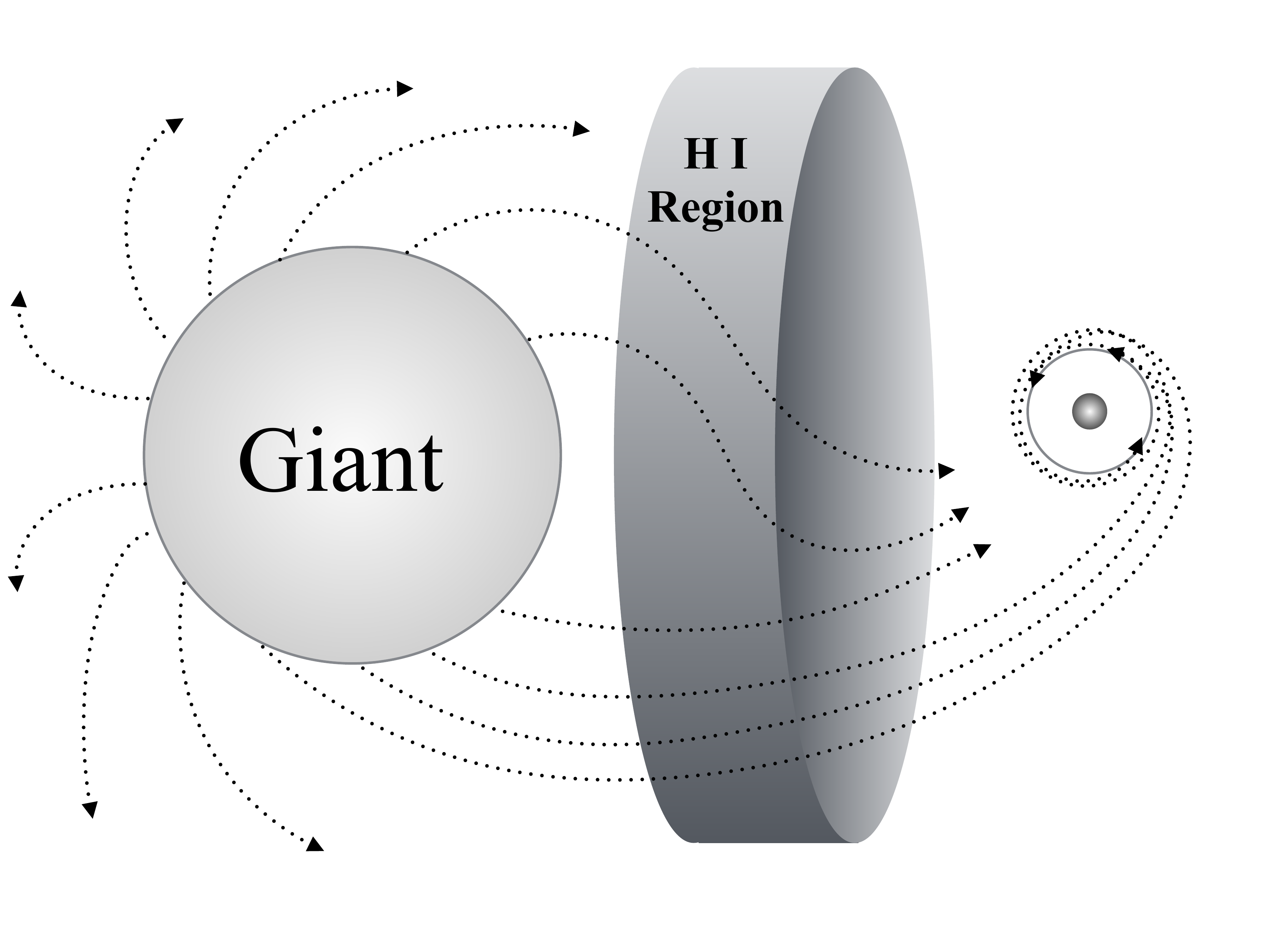}
\caption{A schematic illustration of Raman scattering geometry adopted in this work
for a symbiotic star. We adopt a highly simplified neutral region delineated by a finite cylinder with $R/H=5$
in front of the giant component. The diameter of the neutral cylinder is set to equal the binary separation. The
circular boundary of the  \ion{H}{1} region facing the white dwarf cuts the binary axis 
at its midpoint so that the diameter subtends an angle of $90^\circ$ with respect 
to the white dwarf.
}
\label{symbio}
\end{figure}

\begin{figure*}
\centering
\includegraphics[scale=.17]{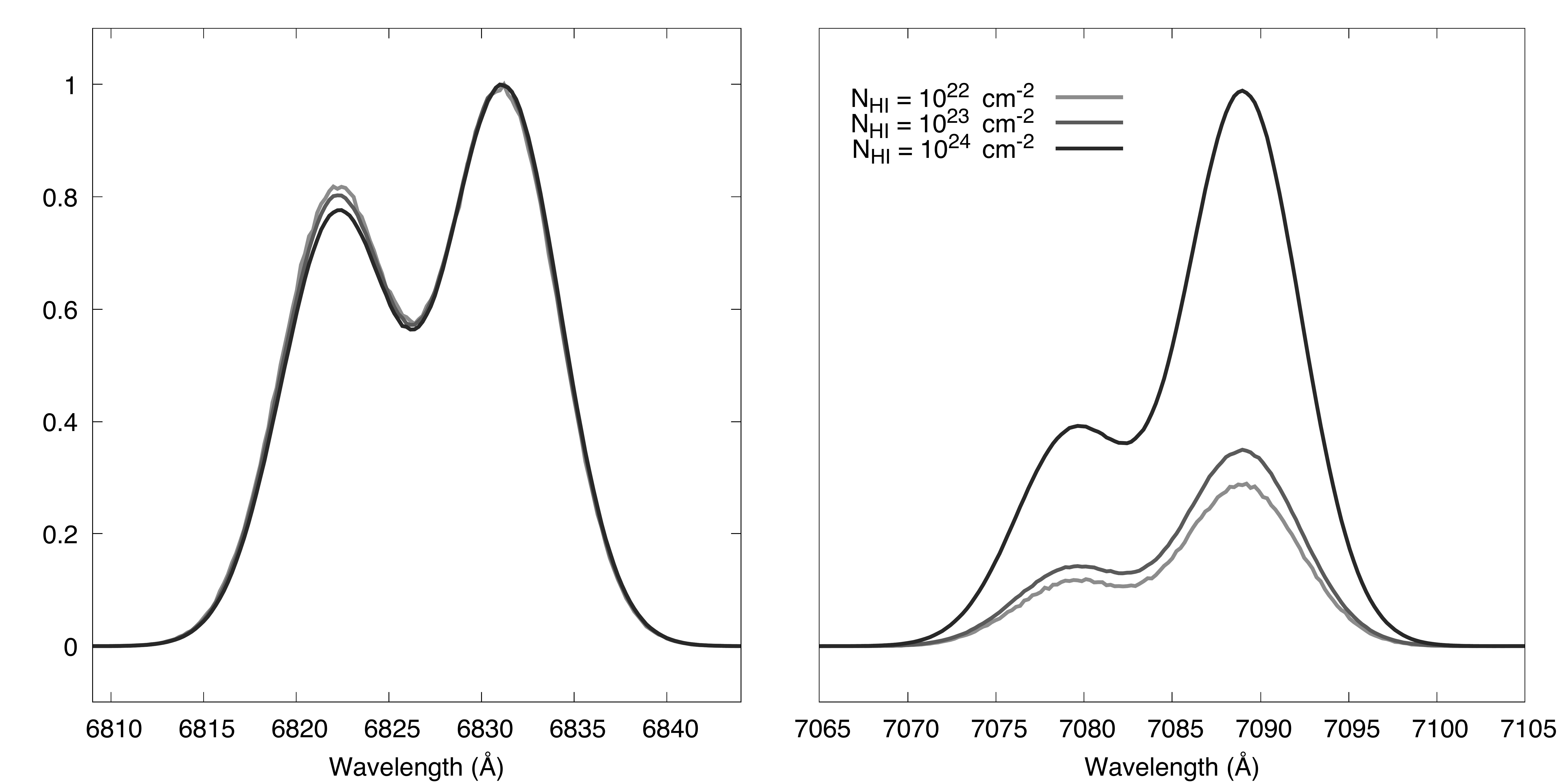}
\caption{Mock spectra of Raman scattered \ion{O}{6}$\lambda\lambda$ 1032 and 1038 formed in in the scattering geometry 
shown in Fig.~\ref{symbio}. The left panel is Raman \ion{O}{6} feature at 6825 \AA\ and the right panel
is for Raman \ion{O}{6} at 7082 \AA.}
\label{mockspectrum}\end{figure*}

In this subsection, we produce  mock spectra of Raman scattered \ion{O}{6} using a  toy model of a
symbiotic star. A schematic illustration of the Raman scattering geometry is shown in Fig.~\ref{symbio}.
A fraction of the slow stellar wind from the giant component is accreted through gravitational
capture by the white dwarf. We place the
neutral region in front of a giant and set up an \ion{O}{6} emission line region around the white dwarf
component.  

A realistic neutral region is expected to take a shape that may be approximated by
a hyperboloid \citep[e.g.][]{taylor84,lee07}. However, in this work we
adopt a much simpler geometry of a finite cylinder in order to focus on the issue of the flux ratio.
In particular, we fix the ratio $R/H=5$ of the cylinder radius $R$ and height $H$.
We  set the diameter of the cylindrical neutral region equal to the binary separation.
In fact, this geometry was considered in Fig.~\ref{fluxratios_res}.

\cite{lee99} proposed that multiple peak structures and  disparity of Raman \ion{O}{6} features can 
be understood if the \ion{O}{6} emission region is a part of the accretion flow around the white
dwarf component. The accretion flow is expected to be convergent on the entering
side, from which the red part of the Raman \ion{O}{6} feature is formed. On the opposite side, the flow 
tends to be divergent partially colliding with the slow stellar wind from the giant. 
In the optically thin region, \ion{O}{6}$\lambda$1032 is expected to be twice stronger than \ion{O}{6}$\lambda$1038 
due to twice larger statistical weight for $s_{1/2}-p_{3/2}$  transition than $s_{1/2}-p_{1/2}$. However, thermalization dictates 
that  \ion{O}{6}$\lambda\lambda$1032 and 
1038 tend to be of similar strength in the optically thick limit \citep[e.g.,][]{schmid99}. Combining this fact 
with the asymmetric accretion flow with the entering side much thicker than the opposite side,
we expect that the blue part of Raman 7082
should be more suppressed compared to its red part than that
of Raman 6825.

\cite{lee07} performed profile analyses of Raman \ion{O}{6} 6825
fluxes for two 'D' type symbiotic stars, V1016~Cygni and HM~Sagittae.
They successfully showed that the profiles are consistent with
the Keplerian motion of the \ion{O}{6} emission region around the
white dwarf component with a typical speed of $\sim 30{\rm\ km\  s^{-1}}$. 
A similar result is also obtained by \cite{heo15}, who investigated
the disparity in the profiles of Raman \ion{O}{6} features. 
We assume that the \ion{O}{6} emission region is
asymmetric in such a way that the entering side of the stellar wind is
significantly denser than the opposite side.  
For the sake of simplicity, it is assumed that
the \ion{O}{6} emission region is perfectly divided into the two regions, that are the red and blue emission
regions. 

Furthermore, we assume that the \ion{O}{6} line profile
viewed from the neutral region is an asymmetric double Gaussian function, where each Gaussian
component describes either the blue or the red emission region. According to \cite{lee07},
the velocity difference of the two peaks in the Raman \ion{O}{6}$\lambda$ 6825 feature of the symbiotic star V1016~Cyg is about 
$60{\rm\ km\ s^{-1}}$. This separation corresponds to the observed width
of $\sim 10$ \AA\, which is typical
for most symbiotic stars exhibiting Raman \ion{O}{6} features. In this work, we set the full width
at half maximum of the Gaussian profile function to be $\Delta v=20{\rm\ km\ s^{-1}}$.  In addition, we
prepare the incident  \ion{O}{6}$\lambda$1032 with the blue peak flux 80 percent weaker than the red peak.
As \cite{heo15} proposed, we simply assume that
the red emission region is characterized by the flux ratio $F(1032)/F(1038)=1$ 
of \ion{O}{6}$\lambda\lambda$ 1032 and 1038
resonance doublet whereas $F(1032)/F(1038)=2$ in the blue emission region. Therefore,
the ratio of the blue and red peaks for \ion{O}{6}$\lambda$1038 is 0.4.

In Fig.~\ref{mockspectrum} we show the mock spectra obtained from our Monte Carlo simulations based on
the simple model illustrated in Fig.~\ref{symbio}.  Here, we obtain Raman \ion{O}{6} photons for various
values of $N_{HI}$ of the neutral region in the shape of a finite cylinder. The H~I column density is measured
along the cylinder axis so that $N_{HI}=n_{HI}H$. 
In this figure, we present the results for 3 different values of
$\log N_{HI}=22, 23$ and 24.
The left and right panels show the profiles of Raman \ion{O}{6} features 
at 6825 \AA\ and 7082 \AA, respectively. The vertical axis is a relative flux density and we
normalize the profile so that the maximum of Raman \ion{O}{6} at 6825 has the unit value.

In the left panel, it is noted that the profiles of Raman 6825 \ion{O}{6} feature differ for different values of $N_{HI}$.
As $N_{HI}$ increases, the blue component tends to be increasingly suppressed relative to the red component.
This slight variation is attributed to the decrease of scattering cross section near 
1032 \AA\ as the wavelength increases. The blue and red peaks occur at $\lambda=
1031.809$ \AA\  and $1032.015$ \AA, respectively, where the relative decrease 
in cross section is  about 6 percent and the relative increase in branching ratio 
is only 0.1 percent. This implies that the decrease in the cross section is mainly responsible 
for the variation of the blue peak relative to the red peak amounting to 4 percent.

In the right panel, we note a significant flux variation in Raman \ion{O}{6} 7082. In the high column density limit,
the red peaks of Raman 6825 and Raman 7082 are comparable. In the opposite limit of low column density
the flux ratio of the red peaks approaches 3.47, the ratio of Raman scattering cross sections.

\subsection{{\it CFHT} Spectra}

\begin{figure*}
\centering
\includegraphics[scale=.19]{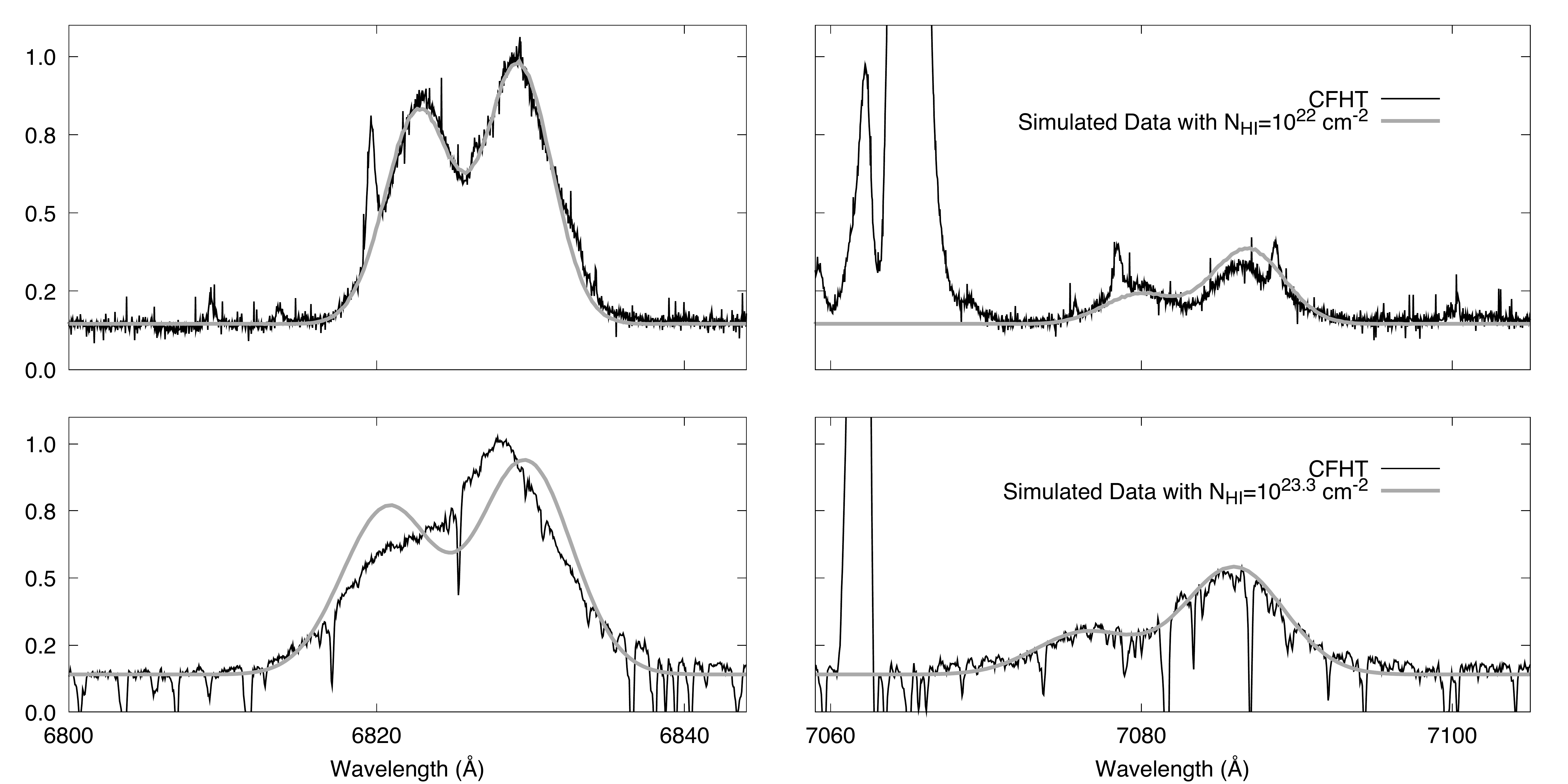}
\caption{{\it CFHT} spectra of Raman scattered \ion{O}{6}$\lambda\lambda$ 1032 and 1038 near 6825 \AA\
and 7082 \AA\ of a 'D' type symbiotic star HM~Sge (top panels) and an 'S' type symbiotic star
AG~Dra (bottom panels). Profiles obtained from our Monte Carlo simulations are overplotted to the {\it CFHT} data.
}
\label{cfht}
\end{figure*}

In Fig.~\ref{cfht}, we show our spectra around Raman \ion{O}{6} features
of two symbiotic stars HM~Sge (top panels) and  AG~Dra (bottom panels) 
obtained with ESPaDOnS installed on the 3.6 m {\it Canada-France-Hawaii Telescope}.
The observation of  HM~Sge and AG~Dra was made on  2014 August  16 and September 6, respectively . 
HM~Sge was observed in spectropolarimetric mode for which the spectral resolution is 68,000. 
In the case of AG~Dra, we choose 'object only' spectroscopic mode for which the spectral resolution is 81,000.
The total integration time for HM~Sge and AG~Dra was 9,600 s and 2,000 s, respectively.  
HM~Sge  is a 'D' type symbiotic star,
having a Mira variable as giant companion, whereas AG~Dra is an 'S' type symbiotic star with an
orbital period of 550 days \citep[e.g.][]{fekel00}.

HM~Sge is known to have erupted as a symbiotic nova in 1975 \citep{dokuchaeva76}. It is also known to exhibit
Raman scattered features of \ion{O}{6} and \ion{He}{2} \citep[e.g.][]{birriel04, lee07}. Based on
their spectropolarimetric monitoring observations, \cite{schmid02} proposed that the orbital period of HM~Sge
amounts to $\sim 50$ years with a caveat that the data quality was insufficient for definite conclusion.  
 AG~Dra is known to be a yellow symbiotic star having an early K type giant as mass donor \citep[e.g.][]{leedjarv16}

In the top panels for HM~Sge, the simulated profiles are obtained  
adopting the same finite cylinder model with $N_{HI}=10^{22}{\rm\ cm^{-2}}$ 
considered in the previous subsection. On the other hand, in the
bottom panels, a finite cylinder model with $N_{HI}=10^{23.3}{\rm\ cm^{-2}}$ is used with the other
conditions the same as in the top panels. 
We obtain relatively poor fit of the Monte Carlo simulated profiles to the {\it CFHT} spectra. However, 
the purpose of the current profile comparison is not a detailed profile fit but a simple estimate 
of the representative neutral column density of the scattering region from the ratio $F(6825)/F(7082)$.
 From this comparison, it is clear that AG~Dra is characterized by a neutral region with $N_{HI}$
 an order of magnitude larger than that of HM~Sge.
We suggest that this is attributed to the closeness of the giant component to the
white dwarf in 'S' type symbiotic stars.

As \cite{schmid99} explained in their observations of a number of symbiotic stars, `S' type symbiotics
systematically show higher Raman conversion efficiency than `D' type symbiotics. Based on near simultaneous
far UV and optical spectroscopic observations, \cite{birriel00} also summarized in their Table~11 higher ratios
$[N_{6825}/N_{1032}]/[N_{7082}/N_{1038}]$ in `S' type symbiotics such as Z~And and AG~Dra 
than in `D' type symbiotic stars V1016~Cyg and RR~Tel.

\section{Discussion}

\subsection{Flux Ratio $F(6825)/F(7082)$}

In this work, we have investigated the flux ratio $F(6825)/F(7082)$  using a Monte Carlo technique mainly as a
function of $N_{HI}$ for a few simple scattering geometries. 
The smallest flux ratio of $F(6825)/F(7082)\sim 0.8$ is obtained when the emission source is outside a neutral slab with infinite lateral
dimensions and Rayleigh-Raman optically thick in the normal direction. In this case, a significant fraction of \ion{O}{6}
photons escape from the neutral region through a few Rayleigh scatterings near the illuminated boundary. On the other hand,
the maximum flux ratio of 3.5 is obtained in the case of a spherical neutral region with a moderate
column density of $N_{HI}\sim 10^{22.7}{\rm\ cm^{-2}}$ inside which the \ion{O}{6} emission source
is embedded. This maximum value is attributed to a nonlinear behavior shown in Raman conversion of \ion{O}{6}$\lambda$1032
whereas \ion{O}{6}$\lambda$1038 remains in the linear domain.

According to \cite{espey95} the flux ratios of $F(1032)/F(6825)\simeq 12$ and $F(1038)/F(7082)\simeq 36$ 
measured for the symbiotic star RR~Tel correspond to our result $F(6825)/F(7082)\simeq 3$ in this work
in which we fix the number flux $F(1032)/F(1038)=1$.
The ratio of Raman conversion efficiencies $\simeq 3$ strongly implies that the neutral scattering region is characterized by
$N_{HI}<10^{23}{\rm\ cm^{-2}}$. 
Another notable point is that the flux ratio in the optically thin limit may be irrelevant to real observations 
because Raman fluxes are very weak. If we require Raman conversion efficiency of $>0.1$, then 
the representative column density of the neutral region around RR~Tel is suggested to be found in the range 
$N_{HI}=10^{22-23}{\rm\ cm^{-2}}$.

Attenuation of far UV radiation due to Rayleigh scattering in symbiotic stars is useful in estimating the neutral
column density of the giant atmosphere \citep[e.g.][]{isliker89,schmutz94, shagatova16}. Along a line of sight 
from the white dwarf to the giant photosphere, it is expected that
$N_{HI}>10^{24}{\rm\ cm^{-2}}$. For example, \cite{shagatova16} proposed a neutral column
density $N_{HI}\sim 10^{25}{\rm\ cm^{-2}}$ toward the giant atmosphere of the symbiotic star
EG~And based on their measurement using {\it IUE} and {\it HST} spectra. From this work, we propose 
that the flux ratio $F(6825)/F(7082)$ can be a rough proxy to estimate $N_{HI}$ in symbiotic stars.

\subsection{\ion{H}{1} Column Density}

With the giant companion being closer to the white dwarf primary, 'S' type symbiotics tend
to be in a more favorable condition to establish a neutral scattering region with significantly higher $N_{HI}$ than
'D' type symbiotics. This leads to the ratio $F(6825)/F(7082)$ systematically larger in 'D' type symbiotics
than in 'S' type symbiotics, which was pointed out by \cite{schmid99}. However, not only the separation between
the binary components, but also very different properties of the wind from cool components in S-type and D-type symbiotics can affect their \ion{H}{1} region.

A crude estimate of the representative $N_{HI}$ can be given by assuming that the mass loss process
is characterized by a spherical stellar wind with velocity field at a radial distance $r$ from the giant
\begin{equation}
v(r) = v_\infty \left( 1-{R_*\over r} \right)^\beta.
\end{equation}
Here, $v_\infty$ is the wind terminal velocity, $\beta$ is a parameter of order unity and $R_*$ is the launching radius of the giant stellar wind \citep[e.g.][]{lamers99}.  A line of sight from the white dwarf is specified by the impact parameter $b$ with
respect to the giant.  If we disregard the photoionization process due to strong far UV radiation from the white dwarf,
the \ion{H}{1} column density along this line is given by
\begin{equation}
N_{HI}={2\dot M\over 4\pi \mu m_p v_\infty} \int_{b}^\infty {dr\over (r-R_*)\sqrt{r^2-b^2}},
\end{equation}
with the choice of $\beta=1$ and $b>R_*$ \citep{nussbaumer87}. Here, $\mu$ is the mean molecular weight of the giant stellar wind.
 Whereas the definite integral is explicitly written as
 \begin{equation}
 N_{HI}={\dot M\over \pi \mu m_p v_\infty\sqrt{b^2-R_*^2}}
 \left[ {\pi\over2}-\tan^{-1}\sqrt{{b-R_*\over b+R_*}} \right],
\end{equation}
a more useful approach is to obtain a rough approximation by taking $\mu$ and the value in the parenthesis to be unity
\citep[e.g.][]{lee07}.
We may take a typical impact parameter $b$ as the binary separation $A$. In the case of `D' type symbiotics $A\gg R_*$, whereas
for `S' type symbiotics the binary separation $b$ is comparable to or a few times $R_*$. With this consideration we may write
\begin{equation}
N_{HI}\sim \left({\dot M_7\over v_{10} A_{14}} \right) 10^{22}{\rm\ cm^{-2}}
\end{equation}
where $\dot M_7=10^{-7}M_\odot {\rm\ yr^{-1}}$, $v_{10}=v_\infty/[10{\rm\ km\ s^{-1}}]$
and $A_{14}=A/10^{14}{\rm\ cm}$.

\cite{seaquist90} proposed that the mass loss rate is correlated with the spectral type in such a way that
later spectral type giants tend to lose mass faster \citep[see also][]{seaquist93}.
 In the case `D' type symbiotic star HM~Sge, our result of
 $N_{HI}\sim 10^{22}{\rm\ cm^{-2}}$ implies a relation $\dot M_7/(v_{10}A_{14})\sim 1$.
Considering that the orbital parameters are only poorly known for most `D' type symbiotic stars, Raman \ion{O}{6}
spectroscopy can be of significant use in putting constraints on the orbital parameters.

\subsection{Mass Loss Rate}

\cite{sekeras15} used Raman scattered \ion{He}{2} feature at 6545 \AA\  to give an estimate of the mass loss
rate of the `D' type symbiotic star V1016~Cygni. They proposed a mass loss rate $\sim 10^{-6}{\rm\ M_\odot\ yr^{-1}}$
based on their analysis. A lower estimate of $3.6\times 10^{-7}{\rm\ M_\odot\ yr^{-1}}$ was proposed by \cite{jung04},
who used the center shift of Raman scattered \ion{He}{2} feature at 4850 \AA\ attributed to the atomic physics
of Raman scattering. Their estimate was $N_{HI}=4\times10^{21}{\rm\
cm^{-2}}$, which is quite small for effective Raman conversion of \ion{O}{6}. Because the line center 
of Raman scattered \ion{He}{2}
feature can also be affected from the kinematics of the \ion{H}{1} region with respect 
to the \ion{He}{2} emission region, there is possibility that
their estimate  would be modified to a higher \ion{H}{1} column density.

Considering that \ion{He}{2}$\lambda\lambda$1025 and 972
have cross sections $\sim 10^{-22}{\rm\ cm^{2}}$, we may expect higher Raman conversion efficiencies
in the formation of Raman \ion{He}{2} features at 6545 \AA\ and 4850 \AA\ than \ion{O}{6} lines. 
For \ion{He}{2}$\lambda$949,
the cross section $\sim 10^{-21}{\rm\ cm^{2}}$ is significantly smaller than the two lower transition lines, 
leading to lower Raman conversion efficiency. One complicating factor in the
analysis of Raman scattering involving \ion{He}{2} is that there are many scattering branches into the IR region.
Sophisticated simulations designed to explain various Raman scattered features including \ion{O}{6} and \ion{He}{2}
will provide very strong constraints on the scattering geometry and mass loss rate involving the giant component.

\subsection{Spectropolarimetric Implications}

The difference in scattering numbers affects the degree of linear polarization 
as well as the flux ratio $F(6825)/F(7082)$. As the scattering number increases, the radiation field tends
to be isotropized resulting in weak polarization. 
\cite{schmid90} reported that the Raman \ion{O}{6} 7082 in the symbiotic star He2-38 exhibits the
degree of linear polarization of 9.2 percent, which is in contrast with the much more weakly 
polarized Raman \ion{O}{6} 6825 with the degree of 5.3 percent. They pointed out that the this difference 
can be understood because Raman \ion{O}{6} 7082 photons are formed with less number of scatterings 
than Raman \ion{O}{6} 6825 photons. As is shown in this work, the mean
scattering number is a complicated function of the scattering optical depth and the branching
ratio.

This work will be extended to obtain Raman fluxes according 
to the emergent wave vector for further polarimetric analysis.
It is expected that combination of these differences in flux and polarization apparent in spectropolarimetric
data will be used to put strong constraints on the mass loss and transfer processes occurring in
symbiotic stars.

\section*{Acknowledgements}

The authors are very grateful to the anonymous referee for constructive comments.
This work was supported by K-GMT Science Program (PID: 14BK002) funded through Korea GMT Project
operated by Korea Astronomy and Space Science Institute.
 This research was also supported by the Korea Astronomy and Space Science 
Institute under the R\&D program(Project No. 2015-1-320-18) supervised 
by the Ministry of Science, ICT and Future Planning.

\clearpage



\begin{thebibliography}{}

\bibitem[\protect\citeauthoryear{Ahn \& Lee}{2015}]{ahn15}
Ahn, S. -H., Lee, H.-W., 2015, Journal of the Korean Astronomical
Society, 48, 195

\bibitem[\protect\citeauthoryear{Allen}{1980}]{allen80}
Allen, D. A., 1980, MNRAS, 190, 75


\bibitem[\protect\citeauthoryear{Angeloni et al.}{2010}]{angeloni10}
Angeloni, R., Contini, M.,  Ciroi, S.,  Rafanelli, P., 2010, MNRAS, 402, 2075

\bibitem[\protect\citeauthoryear{Angeloni et al.}{2011}]{angeloni11}
Angeloni, R., Di Mille, F., Bland-Hawthorn, J., Osip, D. J.,  2011,  ApJL, 743,  L8

\bibitem[\protect\citeauthoryear{Angeloni et al.}{2012}]{angeloni12}
Angeloni, R., Di Mille, F., Ferreira Lopes,  C. E., Masetti, N., 2012,  ApJL, 756, L21


\bibitem[\protect\citeauthoryear{Birriel}{2004}]{birriel04}
Birriel, J. J., 2004, ApJ, 612, 1136


\bibitem[\protect\citeauthoryear{Birriel et al.}{2000}]{birriel00}
Birriel, J. J., Espey, B. R., Schulte-Ladbeck, R. E., 2000, ApJ, 545, 1020

\bibitem[\protect\citeauthoryear{Chang et al.}{2015}]{chang15}
Chang, S.-J., Heo, J.-E., Di Mille, F., Angeloni, R., Palma, T., Lee, H.-W.,
2015, ApJ, 814, 98


\bibitem[\protect\citeauthoryear{Dokuchaeva}{1976}]{dokuchaeva76}
Dokuchaeva, O. D., 1976, IBVS, 1189, 1  

\bibitem[\protect\citeauthoryear{Espey et al.}{1995}]{espey95}
Espey, B. R., Schulte-Ladbeck, R. E, Kriss, G. A., Hamann, F., Schmid, H. M.,  Johnson, J. J., 
1995, ApJL, 454, L61

\bibitem[\protect\citeauthoryear{Fekel et al.}{2000}]{fekel00}
Fekel, F. C., Hinkle, K. H., Joyce, R. R.,  Skrutskie, M. F., 2000,  AJ, 120, 3255

\bibitem[\protect\citeauthoryear{Harries \& Howarth}{1996}]{harries96}
Harries, T. J., \& Howarth, I. D., 1996, A\&AS, 119, 61

\bibitem[\protect\citeauthoryear{Heo \& Lee}{2015}]{heo15}
Heo, J. -E., Lee, H.-W., 2015, Journal of the Korean Astronomical
Society, 48, 105


\bibitem[\protect\citeauthoryear{Isliker et al.}{1989}]{isliker89}
Isliker, H., Nussbaumer, H.,  Vogel, M., 1989, A\&A, 219, 271

\bibitem[\protect\citeauthoryear{Jung \& Lee}{2004}]{jung04}
Jung, Y.-C., \& Lee, H.-W., 2004, MNRAS, 355, 221


\bibitem[\protect\citeauthoryear{Lamers \& Cassinelli}{1999}]{lamers99}
Lamers, H. J. G. L. M., \& Cassinelli, J. P., 1999, Introduction to Stellar Winds
(Cambridge: Cambridge Univ. Press)


\bibitem[\protect\citeauthoryear{Lee \& Kang}{2007}]{lee07}
Lee, H.-W., Kang, S, 2007, ApJ, 669, 1156

\bibitem[\protect\citeauthoryear{Lee \& Lee}{1997a}]{lee97a}
Lee, H.-W., Lee, K. W., 1997, MNRAS, 287, 211


\bibitem[\protect\citeauthoryear{Lee \& Lee}{1997b}]{lee97b}
Lee, K. W., Lee, H. -W., 1997, MNRAS, 292, 573



\bibitem[\protect\citeauthoryear{Lee \& Park}{1999}]{lee99}
Lee, H.-W., \& Park, M.-G., 1999, ApJL, 515, L89


\bibitem[\protect\citeauthoryear{Leedj \"a rv et al.}{2016}]{leedjarv16} 
 Leedj\"arv, L., G\'alis, R., Hric, L., Merc, J., Burmeister, M, 2016, MNRAS, 456, 2558 

\bibitem[\protect\citeauthoryear{Miko\l ajewska}{2012}]{mikolajewska12}
Miko\l ajewska, M., 2012, Baltic Astronomy, 21, 5

\bibitem[\protect\citeauthoryear{Moore}{1979}]{moore79}
Moore, C. E. 1979, NSRDS-NBS 3, sec{8}

\bibitem[\protect\citeauthoryear{Nussbaumer et al.}{1989}]{nussbaumer89}
Nussbaumer, H.,  Schmid, H. M.,  Vogel, M., 1989,  A\&A, 211, L27


\bibitem[\protect\citeauthoryear{Nussbaumer \& Vogel}{1989}]{nussbaumer87}
Nussbaumer,  Vogel, M., 1987,  A\&A, 182, 51


\bibitem[\protect\citeauthoryear{Saslow \& Mills}{1969}]{saslow69}
Saslow, W. M., Mills, D. L., 1969, PhRv, 187, 1025

\bibitem[\protect\citeauthoryear{Schmid}{1989}]{schmid89}
Schmid, H. M., 1989, A\&A, 211, L31

\bibitem[\protect\citeauthoryear{Schmid}{1992}]{schmid92}
Schmid, H. M., 1992, A\&A, 254, 224

\bibitem[\protect\citeauthoryear{Schmid}{1996}]{schmid96}
Schmid, H. M., 1996, MNRAS, 282, 511

\bibitem[\protect\citeauthoryear{Schmid \& Schild}{1990}]{schmid90}
Schmid, H. M., Schild, H. 1990, A\&A, 236, L13

\bibitem[\protect\citeauthoryear{Schmid et al.}{1999}]{schmid99}
Schmid, H. M., et al., 1999, A\&A, 348, 950

\bibitem[\protect\citeauthoryear{Schmid \& Schild}{2002}]{schmid02}
Schmid, H. M., \& Schild, H., 2002, A\&A, 395, 117


\bibitem[\protect\citeauthoryear{Schmutz et al.}{1994}]{schmutz94}
Schmutz, W., Schild, H., M\"urset, U. Schmid, H. M., 1994, A\&A, 288, 819


\bibitem[\protect\citeauthoryear{Seaquist \& Taylor}{1990}]{seaquist90}
Seaquist, E. R.,  Taylor, A. R., 1990, ApJ, 349, 313

\bibitem[\protect\citeauthoryear{Seaquist et al.}{1993}]{seaquist93}
Seaquist, E. R., Krogulec, M., Taylor, A. R., 1993, ApJ, 410, 260


\bibitem[\protect\citeauthoryear{Seker\'a\v{s} \& Skopal}{2015}]{sekeras15}
Seker\'a\v{s}, M.,  \& Skopal,  A.,  2015, ApJ, 812, 162

\bibitem[\protect\citeauthoryear{Shagatova et al.}{2016}]{shagatova16}
Shagatova, N., Skopal, A., Carikov\'a, Z., 2016, A\&A, 588,  A83

\bibitem[\protect\citeauthoryear{Sokoloski \& Bildsten}{2010}]{sokoloski10}
Sokoloski, J. L., Bildsten, L., 2010, ApJ, 723, 1188


\bibitem[\protect\citeauthoryear{Sokoloski et al.}{2001}]{sokoloski01}
Sokoloski, J. L., Bildsten, L., Ho, W. C. G. 2001, MNRAS, 326, 553

\bibitem[\protect\citeauthoryear{Taylor \& Seaquist}{1984}]{taylor84}
Taylor, A. R., \& Seaquist, E. R. 1984, ApJ, 286, 263


\bibitem[\protect\citeauthoryear{Warner}{1995}]{warner95}
Warner, B.. 1995, Catalcysmic Variable Stars, Cambridge Univ. Press, Cambridge

\bibitem[\protect\citeauthoryear{Whitelock}{1987}]{whitelock87}
Whitelock, P. A., 1987, PASP, 99, 573

\bibitem[\protect\citeauthoryear{Zamanov et al.}{2015}]{zamanov15}
Zamanov, R., Latev, G., Boeva, S., Sokoloski, J. L., Stoyanov, K., 
Bachev, R., Spassov, B., Nikolov, G., Golev, V., Ibryamov, S., 2015, MNRAS, 450, 3958


\end{thebibliography}
\end{document}